\documentclass[twocolumn,superscriptaddress,amsmath,amssymb,showpacs,prb]{revtex4-2}

\usepackage{graphicx}
\usepackage{color}
\renewcommand{\phi}{\varphi}

\begin{document}

\title{Structure and motifs of iron oxides from 1 to 3 TPa}

\author{Feng Zheng}
\affiliation{Department of Physics, OSED,\\
Key Laboratory of Low Dimensional Condensed Matter Physics\\ (Department of Education of Fujian Province)\\
Jiujiang Research institute, Xiamen University, Xiamen 361005, China.}
\author{Yang Sun}
\email{ys3339@columbia.edu}
\affiliation{Department of Applied Physics and Applied Mathematics, Columbia University, New York, NY, 10027, USA}
\author{Renhai Wang}
\affiliation{School of Physics and Optoelectronic Engineering, Guangdong University of Technology, Guangzhou 510006, China}
\affiliation{Department of Physics, Iowa State University, Ames, Iowa 50011, United States}
\author{Yimei Fang}
\affiliation{Department of Physics, OSED,\\
Key Laboratory of Low Dimensional Condensed Matter Physics\\ (Department of Education of Fujian Province)\\
Jiujiang Research institute, Xiamen University, Xiamen 361005, China.}
\author{Feng Zhang}
\affiliation{Department of Physics, Iowa State University, Ames, Iowa 50011, United States}
\author{Bo Da}
\affiliation{Research and Services Division of Materials Data and Integrated System, National Institute for Materials Science, 1-1 Namiki, Tsukuba, Ibaraki 305-0044, Japan}

\author{Shunqing Wu}
\email{wsq@xmu.edu.cn}
\affiliation{Department of Physics, OSED,\\
Key Laboratory of Low Dimensional Condensed Matter Physics\\ (Department of Education of Fujian Province)\\
Jiujiang Research institute, Xiamen University, Xiamen 361005, China.}
\author{Cai-Zhuang Wang}
\affiliation{Department of Physics, Iowa State University, Ames, Iowa 50011, United States}
\author{Renata M. Wentzcovitch}
\email{rmw2150@columbia.edu}
\affiliation{Department of Applied Physics and Applied Mathematics, Columbia University, New York, NY, 10027, USA}
\affiliation{Department of Earth and Environmental Sciences, Columbia University, New York, NY, 10027, USA}
\affiliation{Lamont-Doherty Earth Observatory, Columbia University, Palisades, NY, 10964, USA}

\author{Kai-Ming Ho}
\affiliation{Department of Physics, Iowa State University, Ames, Iowa 50011, United States}

\begin{abstract}

Iron oxides are fundamental components of planet-forming materials. Understanding the Fe-O system\textquotesingle s behavior and properties under high pressure can help us identify many new phases and states possible in exoplanetary interiors, especially terrestrial ones. Using the adaptive genetic algorithm (AGA), we investigate the structure of iron oxides for a wide range of stoichiometries (0.25 $\leq$ $\emph{x}_O$ $\leq$ 0.8) at 1, 2, and 3 TPa. Five unreported ground-state structures with Fe$_2$O, FeO, Fe$_3$O$_5$, FeO$_2$, and FeO$_4$ compositions are identified. 
Phonon calculations confirm their dynamical stability. 
The \textit{ab initio} molecular dynamics simulations confirm the thermal stability of Fe-rich phases at high temperatures. 
The calculated density of states (DOS) suggests that, except for FeO$_4$, all phases are metallic, but their carrier densities decrease with increasing pressure and oxygen content. The cluster alignment analysis of stable and metastable phases shows that several motifs may co-exist in a structure of iron oxides with low O content. In contrast, most iron oxides with high O content adopt a simple BCC motif at TPa pressures. Our results provide a crystal structure database of iron oxides for modeling and understanding the interiors of exoplanets.

\end{abstract}

\date{Apr. 10, 2022}

\maketitle

\section{Introduction}
 Iron and oxygen are the two most significant elements of Earth-like exoplanets~\cite{1doyle2019oxygen}. Studies on iron oxides\textquotesingle~ structures and fundamental properties provide a valuable understanding of exoplanet forming phases, particularly terrestrial ones. Their high-pressure behavior has received considerable attention to advance understanding of Earth\textquotesingle s interior. So far, at ambient or low pressures, three basic iron oxides have been known, i.e., w\"{u}stite FeO~\cite{2fei1994situ} magnetite Fe$_3$O$_4$~\cite{3pasternak1994high,4fei1999situ}, and hematite Fe$_2$O$_3$~\cite{5rozenberg2002high,6badro2002nature}.Previous studies revealed that these three iron oxides undergo complex electronic~\cite{7shim2009electronic}, magnetic~\cite{7shim2009electronic,8ju2012pressure}, and structural transformations~\cite{8ju2012pressure,9ozawa2011phase,10bykova2016structural} at high pressure, which can not only lead to seismic anomalies but also affect geochemical processes in Earth\textquotesingle s interior. Besides the three basic iron oxides, several new stoichiometries of compounds were also synthesized at high pressure, such as Fe$_4$O$_5$~\cite{11lavina2011discovery}, Fe$_5$O$_6$~\cite{12lavina2015unraveling}, Fe$_5$O$_7$~\cite{10bykova2016structural} and Fe$_7$O$_9$~\cite{13sinmyo2016discovery}, suggesting a complex phase diagram of iron oxides. Recently, using ab initio random structure searching (AIRSS), Weerasinghe \emph{et al.} identified a series of stable and metastable Fe-O compounds at 100, 350, and 500 GPa~\cite{14weerasinghe2015computational}, which further broadens the database of the Fe-O system at high pressure. It is worth noting that the predicted pyrite-type FeO$_2$ was later confirmed by experimental synthesis~\cite{15hu2016feo}. This successful discovery is impactful and demonstrates that computational predictions can play a significant role in discovering high-pressure phases.

However, up to now, most attention has been focused on elucidating the nature of structures and phase transitions of iron oxides below 500 GPa. A legitimate question is: what are the subsequent high-pressure phases of iron oxides? The answer can provide insights into the types of coordination preferred by iron in planet-forming silicates and oxides and possible, stable phases in solid parts of terrestrial planetary cores where pressures can reach $\sim$4 TPa ~\cite{16van2019mass}. These exoplanets frequently referred to as "super-Earths"~\cite{17seager2007mass}, have a similar interior structure and composition with Earth, which is dominated by the elements Fe and O ~\cite{1doyle2019oxygen,17seager2007mass}. Modeling and understanding these planetary interiors can help us investigate their potential habitability, but it requires a basic knowledge of planet-forming phases and their properties under extreme conditions.
In this paper, using an adaptive genetic algorithm (AGA)~\cite{18wu2013adaptive}, we study the Fe-O compounds at 1, 2, and 3 TPa across a wide range of stoichiometries (0.25 $\leq$ $\emph{x}_O$ $\leq$ 0.8). The phase stability and electronic properties of five ground-state Fe-O structures are investigated. The local packing motifs in these stable and metastable Fe-O compounds are also analyzed as a function of O contents.

\section{Computational Methods}
In this work, Fe-O\textquotesingle s crystal structures were determined using the AGA method~\cite{18wu2013adaptive}, which combines fast structure exploration by auxiliary classical potentials and the accurate ab initio calculations adaptively and iteratively. The Fe and O atoms\textquotesingle~ initial atomic positions were randomly generated in the GA-loop without assuming the Bravais lattice type, symmetry, atom basis, or unit cell dimensions. The total structure pool in our GA search was set to be 128. Structure searches with auxiliary interatomic potentials were performed 500 consecutive GA generations. Then, the 16 lowest-enthalpy structures at the end of each GA search were selected for single point DFT calculations according to the AGA procedure~\cite{18wu2013adaptive}, whose energies, force, and stress are used to adjust the interatomic potential parameters for the next iteration of GA search. A total of 40 adaptive iterations were performed to obtain the final structures.

Here, the embedded-atom method (EAM)~\cite{19foiles1986embedded} was used as classical auxiliary potential. In EAM, the total energy of an N-atom system was evaluated by
\begin{equation}
E_{total}=\frac{1}{2}\sum\nolimits_{i,j(i\ne j)}^{N}\phi(r_{ij})+\sum\nolimits_{i}F_{i}(n_i)
\end{equation}
where $\phi(r_{ij})$ denotes the pair repulsion between atoms $i$ and $j$ with a distance of $r_{ij}$, $F_{i}(n_i)$ is the embedded term with electron density term $n_i=\sum\nolimits_{j \ne i}\rho_{j}(r_{ij})$ at the site occupied by atom $i$. The fitting parameters in the EAM formula were chosen as follows: The parameters for Fe-Fe interactions were taken from the literature~\cite{20zhou2004misfit}, while the Fe-O and O-O interactions were modeled by Morse function,
\begin{equation}
\phi(r_{ij}) = D[e^{-2\alpha(r_{ij}-r_0)}-2e^{-\alpha(r_{ij}-r_0)}],
\end{equation}
where $D$, $\alpha$ and $r_0$ are the fitting parameters. The density function for O atoms is modeled by an exponentially decaying function
\begin{equation}
\rho(r_{ij}) = \alpha exp[-\beta(r_{ij}-r_{0})],
\end{equation}
$\alpha$ and $\beta$ are fitting parameters. The form proposed by Benerjea and Smith in Ref.~\cite{21banerjea1988origins} was used as the embedding function with fitting parameters $F_0$,$\gamma$ as,

\begin{equation}
F(n) = F_0[1-\gamma {\rm ln} n]n^{\gamma}.
\end{equation}
For Fe, the density function and embedding function parameters were taken from Ref.~\cite{20zhou2004misfit}. The potential fitting was performed by the force-matching method with a stochastic simulated annealing algorithm as implemented in the POTFIT code~\cite{22brommer2006effective,23brommer2007potfit}.

First-principles calculations were carried out using the Quantum ESPRESSO (QE) code~\cite{24giannozzi2009quantum,25giannozzi2017advanced}. Our calculations suggest that the spin-polarized state is unstable at ultrahigh pressures for Fe-O system. As shown in Fig. S1, Fe${_x}$O${_y}$ compounds with finite magnetic moments always shows much higher enthalpies than the non-magnetic solution. This can be attributed to the enhancement of overlap of the atomic wave functions at ultrahigh pressures, which broadens the Fe $d$ bands and destroys the magnetic order.
For this reason, the non-spin-polarized generalized-gradient approximation (GGA) parameterized by Perdew-Burke-Ernzerhof formula (PBE) was used to describe the exchange-correlation energy. 
The pseudopotentials for Fe and O were generated by Vanderbilt's method \cite{a26} with the valence electronic configuration of $3s^23p^63d^{6.5}4s^1$ and $2s^22p^4$, respectively. These potentials were tested and previously used in a few studies \cite{a27} at terapascal pressure, and showed consistent results with the all-electron full-potential calculations for Fe-O phases at 1-3 TPa. 
A kinetic-energy cutoff of 50 Ry for wave functions and 500 Ry for potentials were used. Brillouin-zone integration was performed over a k-point grid of $2\pi \times 0.03~\text{\AA}^{-1}$ in the structure refinement. The convergence thresholds are 0.01 eV/$\text{\AA}$~ for the atomic force, 0.5 kbar for the pressure, and $1 \times 10^{-5}$ eV for the total energy. The structural optimization was performed under constant pressure using the Broydon-Fletcher Goldfarb-Shanno (BFGS) algorithm~\cite{26broyden1970convergence1,27broyden1970convergence2,28fletcher1970new,29goldfarb1970family,30shanno1970conditioning} with variable cell shape. Phonon dispersions were calculated using the finite displacement method as implemented in the PHONOPY software ~\cite{31togo2008first,32togo2015first}.

The \textit{ab initio} molecular dynamics (AIMD) simulations were performed in the canonical (NVT) ensemble using the Nosé-hoover thermostat \cite{a35} as implemented in the Vienna ab initio simulation package \cite{a36,a37}. A plane-wave energy cutoff of 450 eV and gamma-only k-point grids were used. The MD timestep is 1 $fs$. All the AIMD simulations start with the perfect crystals and run for more than 10 $ps$. We used $4\times4\times2$ supercell (192 atoms) for Fe$_2$O, $2\times2\times3$ supercell (192 atoms) for Fe$_3$O$_5$, $4\times4\times4$ supercell (128 atoms) for FeO, $3\times3\times2$ supercell (216 atoms) for FeO$_2$ and $3\times3\times3$ supercell (270 atoms) for FeO$_4$ in the AIMD simulations.

\begin{figure}
\begin{center}
\includegraphics[width=3.5in]{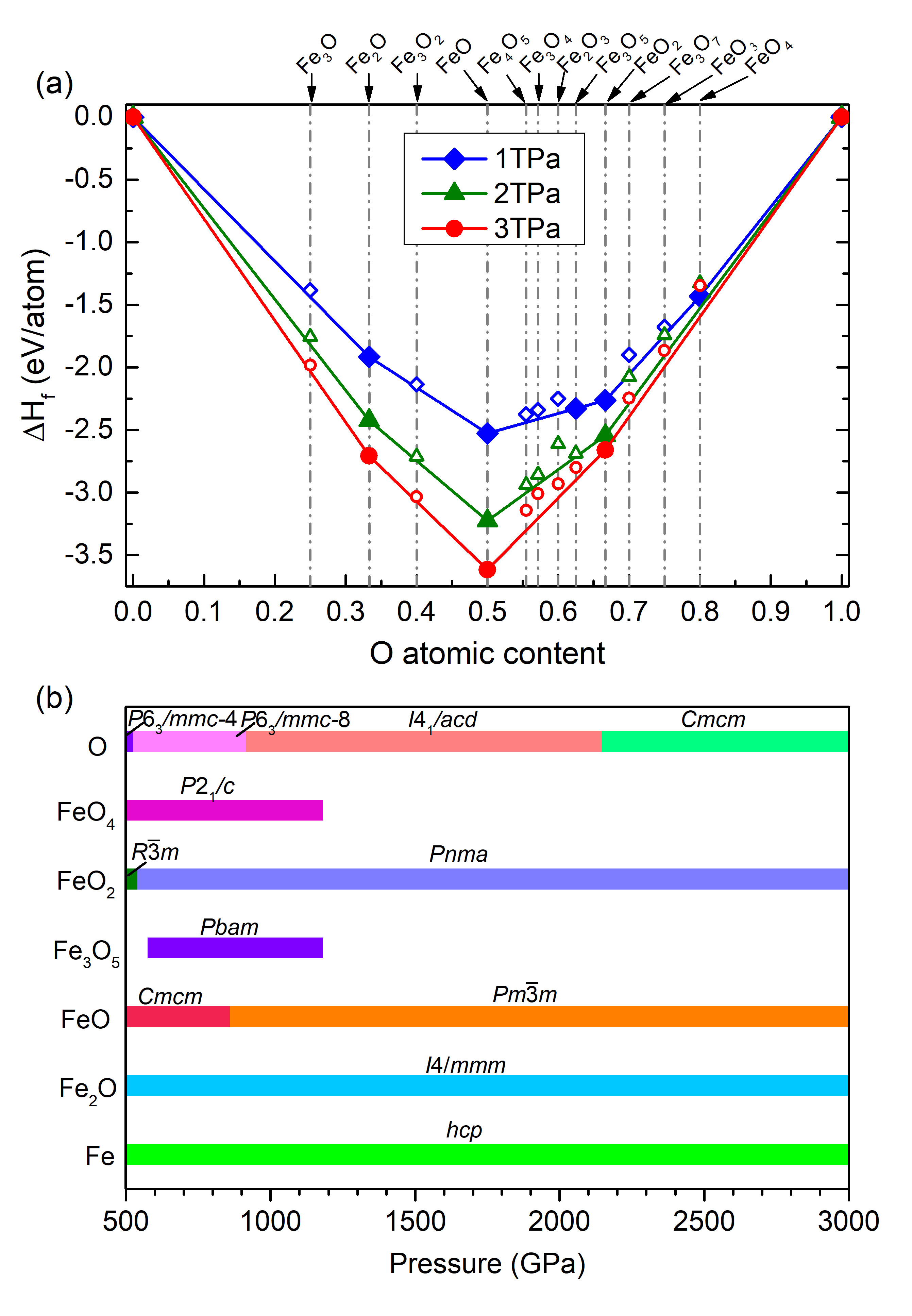}
\end{center}
\caption{(a) Convex hull diagrams of the Fe-O system at 1, 2, and 3 TPa. (b) Pressure stability fields of stable phases in the Fe-O system.}	
\end{figure}

\section{RESULTS AND DISCUSSION }

\subsection{AGA search for the Fe-O system}
In order to obtain low-enthalpy structures of iron oxides, a wide range of stoichiometries of Fe$_x$O$_y$ ($x:y$ = 3:1, 2:1, 3:2, 1:1, 4:5, 3:4, 2:3, 3:5, 1:2, 3:7, 1:3, 1:4) with different formula units (i.e., 1, 2, 3, 4, 5, 6 and 8 f.u.) containing up to 40 atoms are searched at 1, 2 and 3 TPa, respectively. The relative stability of these predicted Fe-O compounds was investigated under the corresponding pressure, depending on the calculated formation enthalpies,
\begin{equation}
H_f = \frac{H_{Fe_{x}O_{y}}-xH_{Fe}-yH_{O}}{x+y},
\end{equation}
where $H$ is the calculated enthalpy for a given structure, $x$ and $y$ are the numbers of atoms of Fe and O, respectively. Before we discuss the stable structures of iron oxides, the crystal structures of pure Fe and O should be clarified. For elemental Fe, our calculated results suggest that the Fe-hcp with $P6_3/mmc$ symmetry is the ground state phase from 1 to 3 TPa. While for oxygen, the $I4_1/acd$ structure \cite{a38} is predicted to be stable at 1 and 2 TPa. At 3 TPa, oxygen adopts a structure with $Cmcm$ symmetry \cite{a38}. The stable structures of Fe and oxygen are shown in Fig. S1. Fig. 1(a) presents convex hulls of the Fe-O system at 1, 2, and 3 TPa. Five Fe$_x$O$_y$ stoichiometries are found in these hulls, i.e., Fe$_2$O, FeO, Fe$_3$O$_5$, FeO$_2$ and FeO$_4$. 
\begin{figure}
\begin{center}
\includegraphics[width=3.5in]{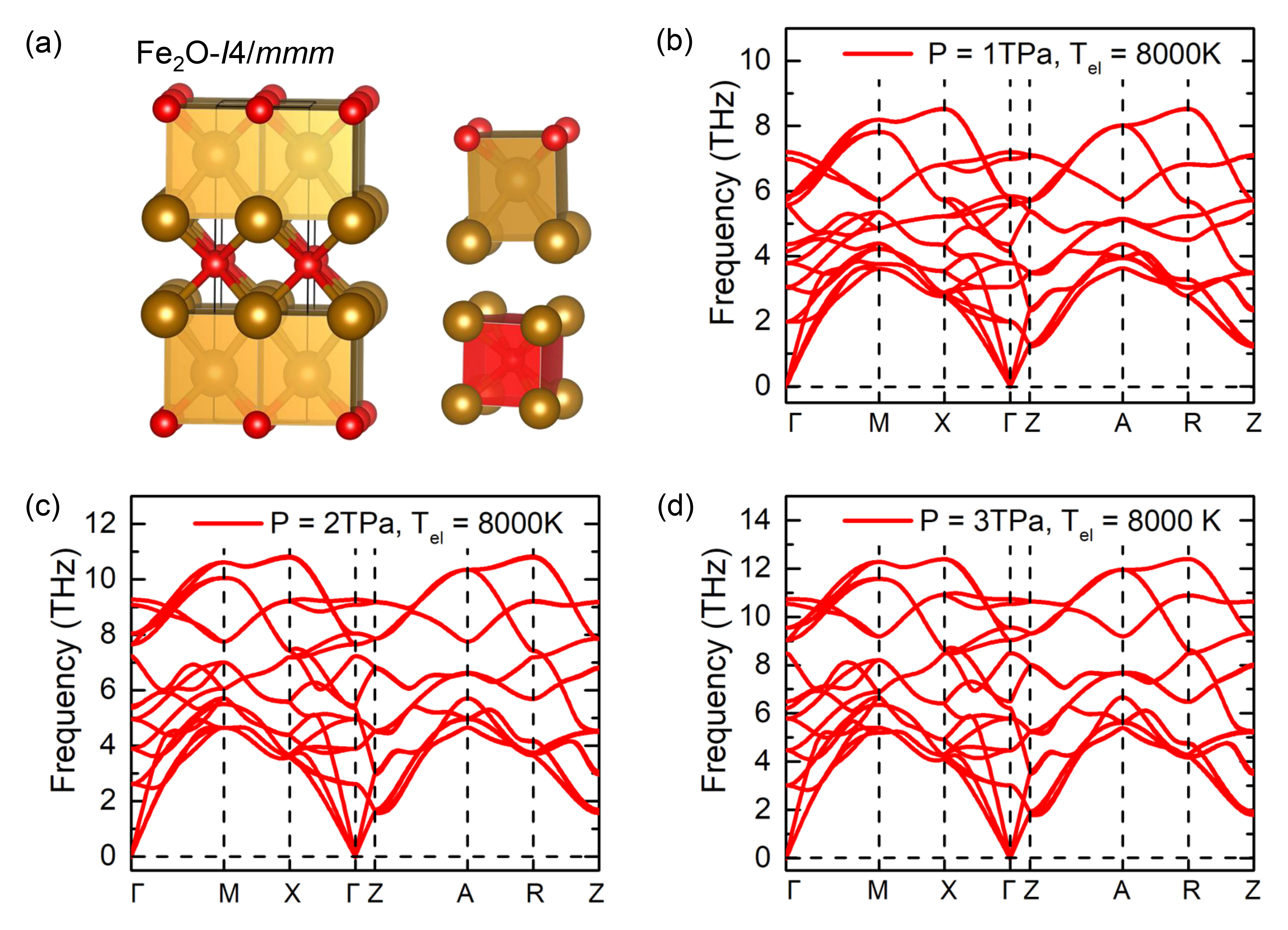}
\end{center}
\caption{(a) Crystal structure and Fe and O coordination polyhedra of $I4/mmm$ Fe$_2$O. Fe and O atoms are denoted by light brown and red spheres, respectively; Phonon dispersions of I4/mmm Fe$_2$O with $T_{el}$ = 8000 K at (b) 1TPa, (c) 2TPa, and (d) 3TPa.}	
\end{figure}
As shown in Fig. 1(b), the stability fields of these phases were investigated from 500 GPa to the upper limit for the pressure considered in Ref.~\cite{14weerasinghe2015computational}, to 3 TPa. For Fe$_2$O, Weerasinghe et al. predicted that an $I4/mmm$ phase could be stable from 288 GPa to 500 GPa~\cite{14weerasinghe2015computational}. Here, we show that it can withstand high pressures up to 3 TPa. Previous DFT calculations show that FeO undergoes a complex structural transformation in the pressure range of the Earth's interior~\cite{14weerasinghe2015computational,33sun2020lda+}. At ultrahigh pressures, our results suggest that the phase with $Cmcm$ symmetry is the ground-state from 500 to 860 GPa. At 860 GPa, the $Cmcm$ phase is predicted to transform into a phase of $Pm\bar{3}m$ symmetry (CsCl-type structure), which remains stable up to 3 TPa. Above 575 GPa, the Fe$_3$O$_5$ phase is stable in an orthorhombic structure with the $Pbam$ symmetry. While, at pressures above 1180 GPa, this phase decomposes into FeO and FeO$_2$. For FeO$_2$, Weerasinghe et al. identified the FeO$_2$ phase with $Pa\bar{3}$ symmetry~\cite{14weerasinghe2015computational}, which is stable from 100 to 456 GPa. This pyrite-type FeO$_2$ phase has recently been confirmed by experiments~\cite{15hu2016feo}. At 456 GPa, the $Pa\bar{3}$ phase is predicted to transform to a phase with $R\bar{3}m$ symmetry~\cite{14weerasinghe2015computational}. Here, we show that $R\bar{3}m$ FeO$_2$ should transform to a new phase with $Pnma$ symmetry at 540 GPa, and $Pnma$ FeO$_2$ can be stable to at least 3 TPa. At 500 GPa, FeO$_4$ adopts a structure with $P2_1/c$ symmetry. At 1180 GPa, $P2_1/c$ FeO$_4$ decomposes into FeO$_2$ and O. The structural parameters of these stable iron oxides are listed in Supplementary Table S1.

\subsection{Crystal structure for stable Fe-O compounds}
\textbf{Fe$_2$O.} Fig. 2 shows the crystal structure and phonon dispersion for tetragonal Fe$_2$O with $I4/mmm$ symmetry. In this structure, each Fe is coordinated to four Fe\textquotesingle s and four O\textquotesingle s, while each O is coordinated to eight Fe\textquotesingle s. These motifs pack in the face-sharing arrangement. This structure is the same as the $I4/mmm$-type phases of Fe$_2$Mg~\cite{34gao2019iron} and Al$_2$S~\cite{35shao2020exotically}. The calculated phonon spectrum confirms that this phase is dynamically stable at 1, 2 and 3 TPa with an electron temperature ($T_{el}$) of 8000 K as seen in Fig. 2 (b)-(d). Because the temperature at the core-mantle boundary of a super-Earth falls within the range from 4000K to 10000 K~\cite{16van2019mass}, the choice of $T_{el}$ = 8000 K is reasonable. Nevertheless, phonon dispersions with $T_{el}$ = 150 K are also presented in Fig. S2, showing no imaginary frequencies in the entire Brillouin zone.

\textbf{FeO.} The phase with $Cmcm$ symmetry is the ground state structure of FeO from 500 to 860 GPa (Fig. S1(d)). From 860 GPa to 3 TPa, FeO stabilizes in the CsCl-type (B2) structure with $Pm\bar{3}m$ symmetry as shown in Fig. 3(a). Phonon calculations confirm its dynamic stability at 1, 2 and 3 TPa with $T_{el}$ = 8000 K (Fig. 3(b)-(d)). At low electron temperature ($T_{el}$ = 150 K), our calculated results show that it is also dynamically stable, as seen in Fig. S3.

\begin{figure}
\begin{center}
\includegraphics[width=3.5in]{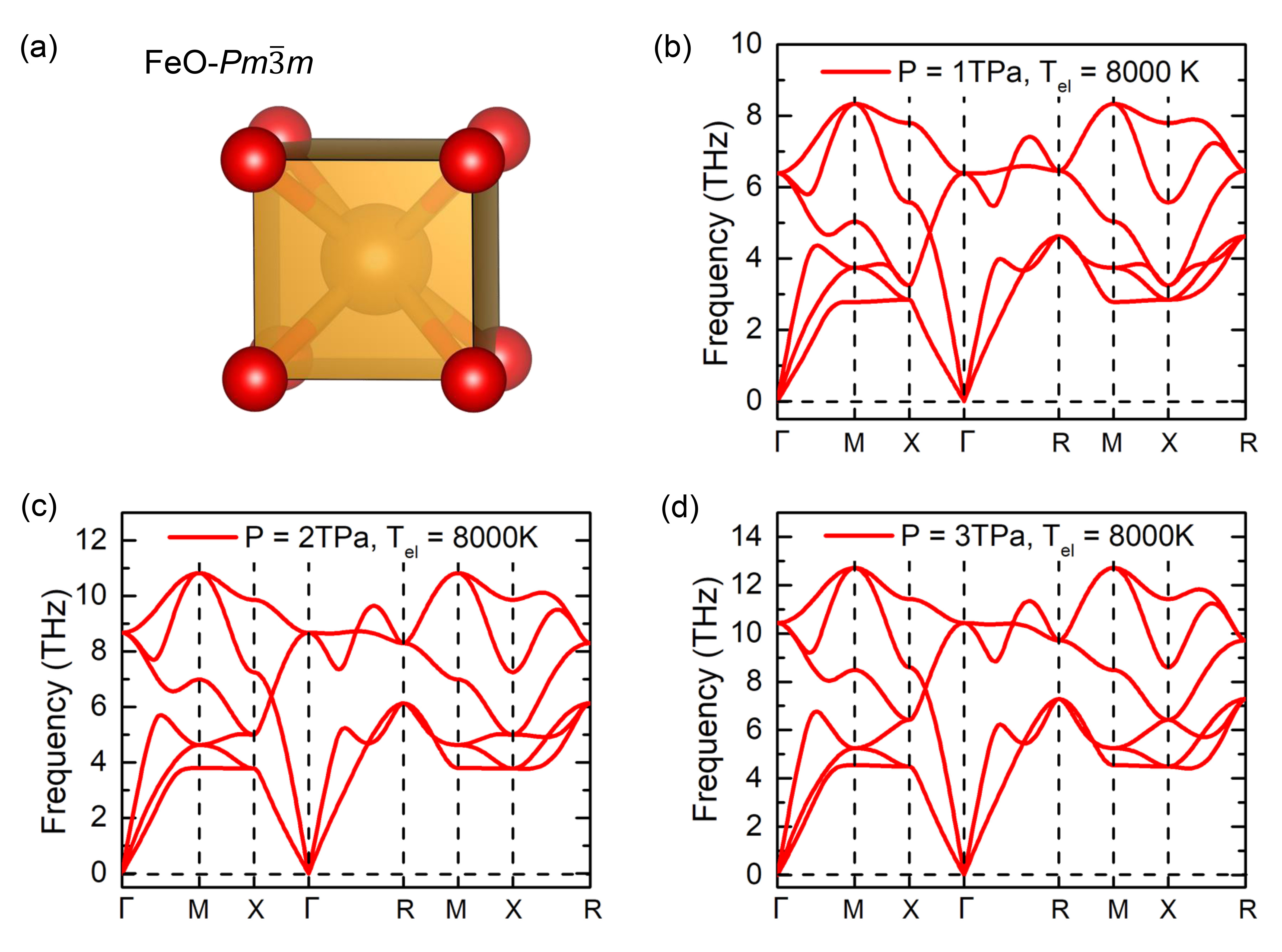}
\end{center}
\caption{(a) Crystal structure of $Pm\bar{3}m$ FeO. Fe and O atoms are denoted by light brown and red spheres, respectively. Phonon dispersions of $Pm\bar{3}m$ FeO with $T_{el}$ = 8000 K at (b) 1TPa, (c) 2TPa and (d) 3TPa.}	
\end{figure}

\textbf{Fe$_3$O$_5$.} From 575 GPa to 1180 GPa, Fe$_3$O$_5$ adopts an orthorhombic phase with $Pbam$ space group (see Fig. 1(b)). In this structure, each Fe is 8-fold coordinated by O\textquotesingle s and form face-shared and edge/face diagonal-shared (an edge in one cube shares with a face diagonal of another cube) cubes, as seen in Fig. 4(a). The calculated phonon dispersion shows the $Pbam$ Fe$_3$O$_5$ is dynamically stable at 1TPa with $T_{el}$ = 8000 K (Fig. 4(b)) and $T_{el}$ = 150 K (Fig. S4). At 1180 GPa, this structure decomposes into FeO and FeO$_2$.
\begin{figure}
\begin{center}
\includegraphics[width=3.5in]{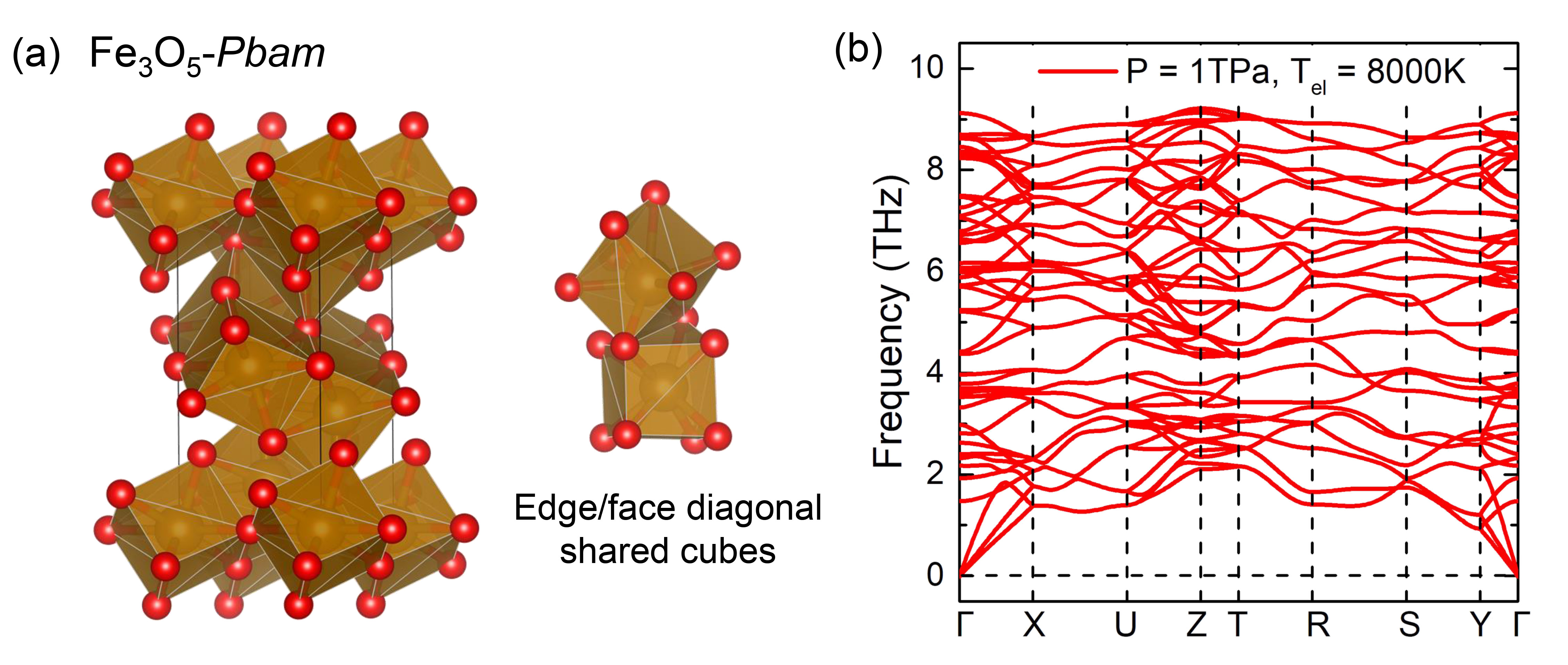}
\end{center}
\caption{(a) Crystal structure and edge/face diagonal shared cubes of $Pnam$ Fe$_3$O$_5$. Light brown and red spheres denote Fe and O, respectively; (b) Phonon dispersion of $Pnam$ Fe$_3$O$_5$ at 1 TPa with $T_{el}$ = 8000 K.  }	
\end{figure}

\textbf{FeO$_2$.} The ground-state structure of FeO$_2$ is orthorhombic with $Pnma$ symmetry from 540 GPa to 3 TPa, as seen in Fig. 5(a). In this structure, each Fe is coordinated by 8 O's forming distorted FeO$_8$ cubes. These cubes pack in a similar arrangement to that in $Pbam$ Fe$_3$O$_5$. The dynamic stability of the $Pnma$ FeO$_2$ is verified by the absence of imaginary frequencies in the phonon dispersion at 1, 2 and 3 TPa with $T_{el}$ = 8000 K as shown in Fig. 5(b)-(d). Phonon dispersions with $T_{el}$ = 150 K are shown in Fig. S5.

\begin{figure}
\begin{center}
\includegraphics[width=3.5in]{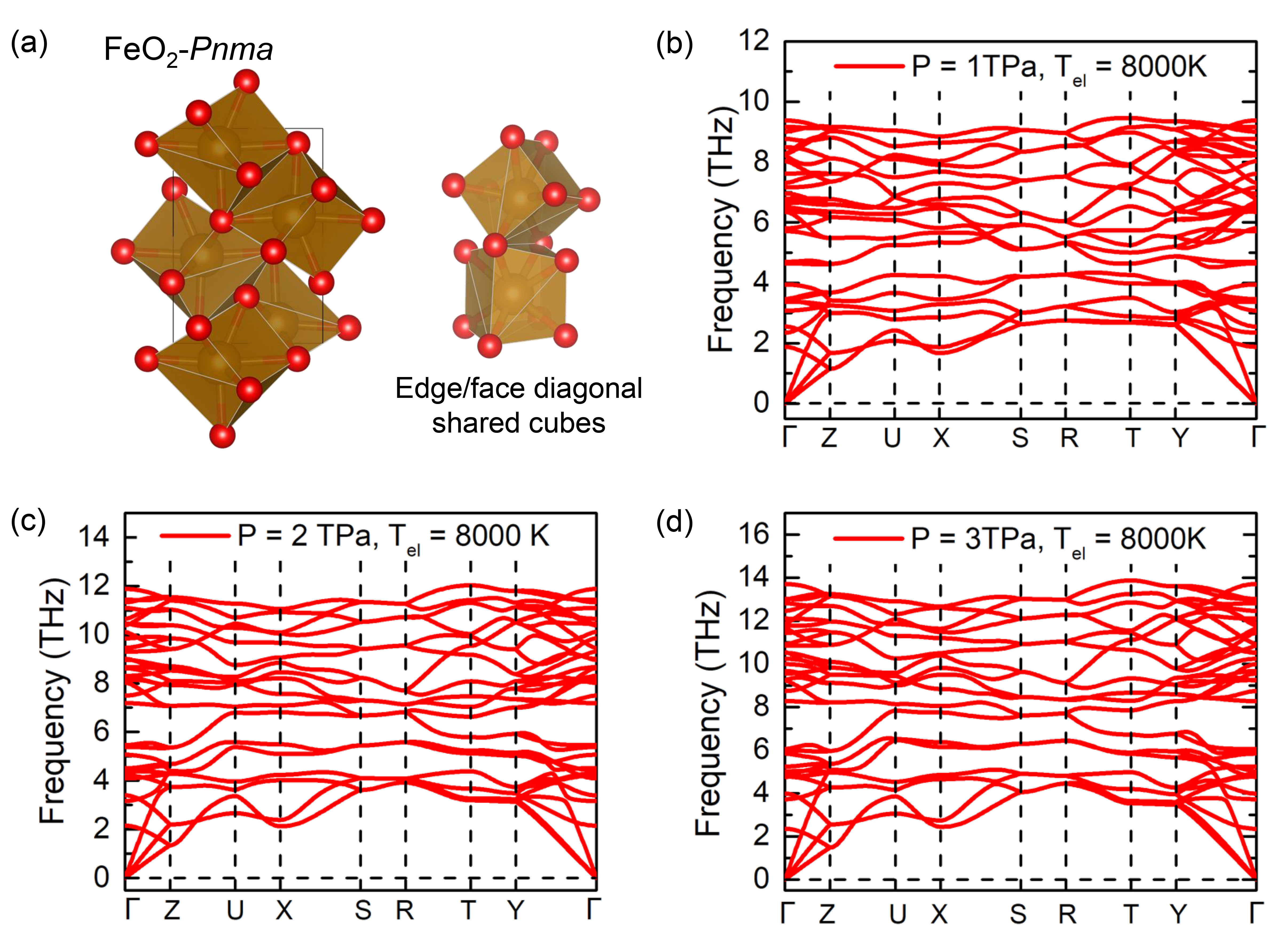}
\end{center}
\caption{(a) Crystal structure and edge/face diagonal shared cubes of $Pnma$ FeO$_2$. Light brown and red spheres denote Fe and O, respectively. Phonon dispersions of $Pnam$ FeO$_2$ at (b) 1 TPa, (c) 2TPa and (d) 3TPa with $T_{el}$ = 8000 K.   }	
\end{figure}

\textbf{FeO$_4$.}
Fig. 6 plots the crystal structure and phonon dispersion for the FeO$_4$ with $P2_1/c$ symmetry. Each Fe is coordinated with 8 O's to form edge-shared cubes. The calculated phonon dispersion shows this $P2_1/c$ FeO$_4$ phase is dynamically stable at 1 TPa with $T_{el}$ = 8000 K (Fig. 6(b)) and $T_{el}$ = 150 K (Fig. S6).

\begin{figure}
\begin{center}
\includegraphics[width=3.5in]{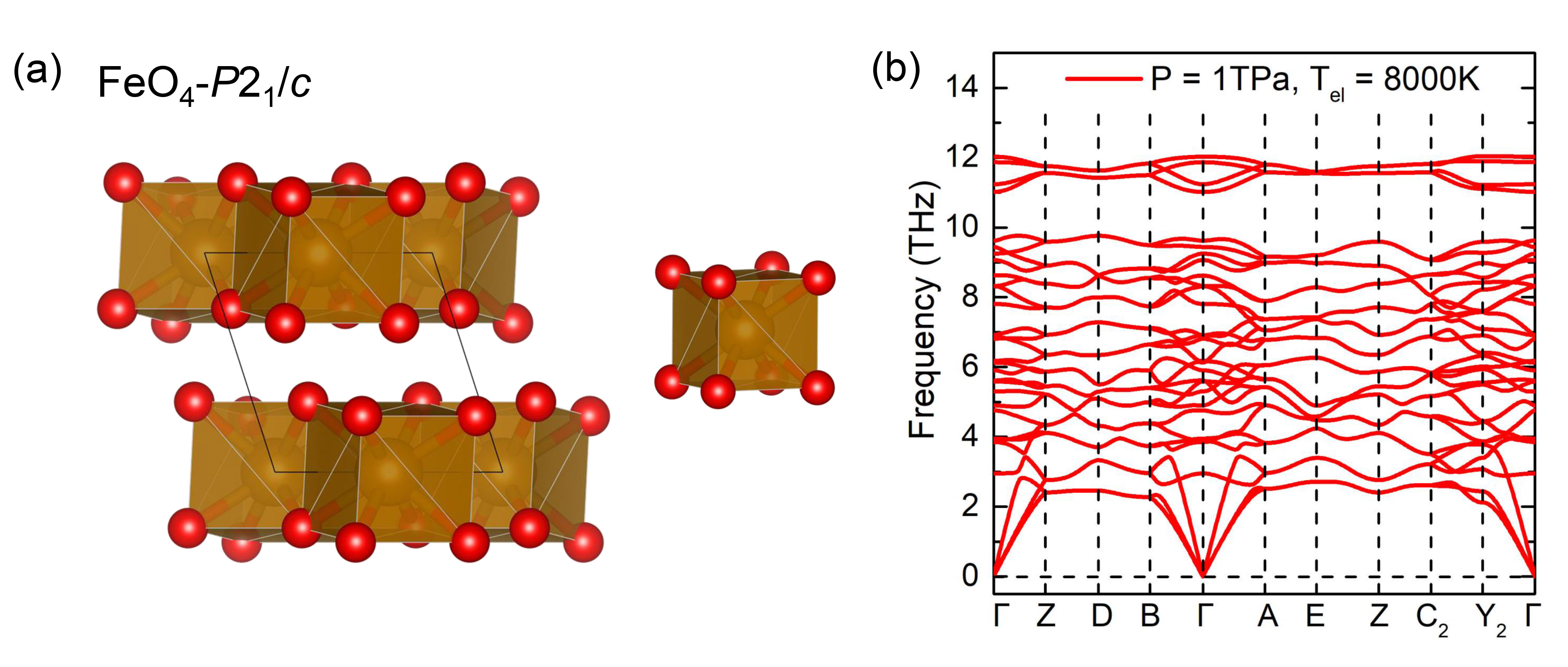}
\end{center}
\caption{(a) Crystal structure of $P2_1/c$ FeO$_4$. Light brown and red spheres denote Fe and O, respectively; (b) Phonon dispersion of $P2_1/c$ FeO$_4$ at 1TPa with $T_{el}$ = 8000 K. }	
\end{figure}

We performed the AIMD simulation to study the thermodynamical stability against melting for the five Fe-O ground states. Both ion and electronic temperatures are 8000 K. This high temperature is close to the estimation of the Super-Earth's interior \cite{16van2019mass}. The mean square displacement (MSD) and superposed atomic positions are shown in Fig. 7, which clearly describes the states in the simulation. During the AIMD, no melting was observed in Fe$_2$O and FeO from ~1-3 TPa. Therefore, the melting points of Fe$_2$O and FeO should be higher than 8000 K under these pressures. Fe$_3$O$_5$ does not show melting at $\sim$ 1 TPa, either. When O content increases beyond Fe$_3$O$_5$, the crystals start to melt. FeO$_2$ and FeO$_4$ both melt at 8000 K and $\sim$ 1 TPa. But when pressure increased to 3 TPa, no melt was observed for FeO$_2$. This is consistent with the general trend that the melting temperature increases with the increasing pressures, as shown in both Fe and FeO P-T diagram under low pressures \cite{9ozawa2011phase,a42}. At $\sim$ 2 TPa, the FeO$_2$ does not melt in the simulation. However, its MSD shows a strong fluctuation. Therefore, the melting point of FeO$_2$ at $\sim$ 2 TPa should be close to 8000 K. These AIMD simulations demonstrate the Fe-poor phases (i.e. FeO$_2$ and FeO$_4$) show lower melting points than the Fe-rich phases (Fe$_2$O and FeO). The Fe-rich phases can have very high melting points, larger than 8000 K at 1-3TPa. Therefore, these Fe-rich crystal phases can be stable under the conditions of Super-Earth's interiors.

\begin{figure}[t]
\begin{center}
\includegraphics[width=3.5in]{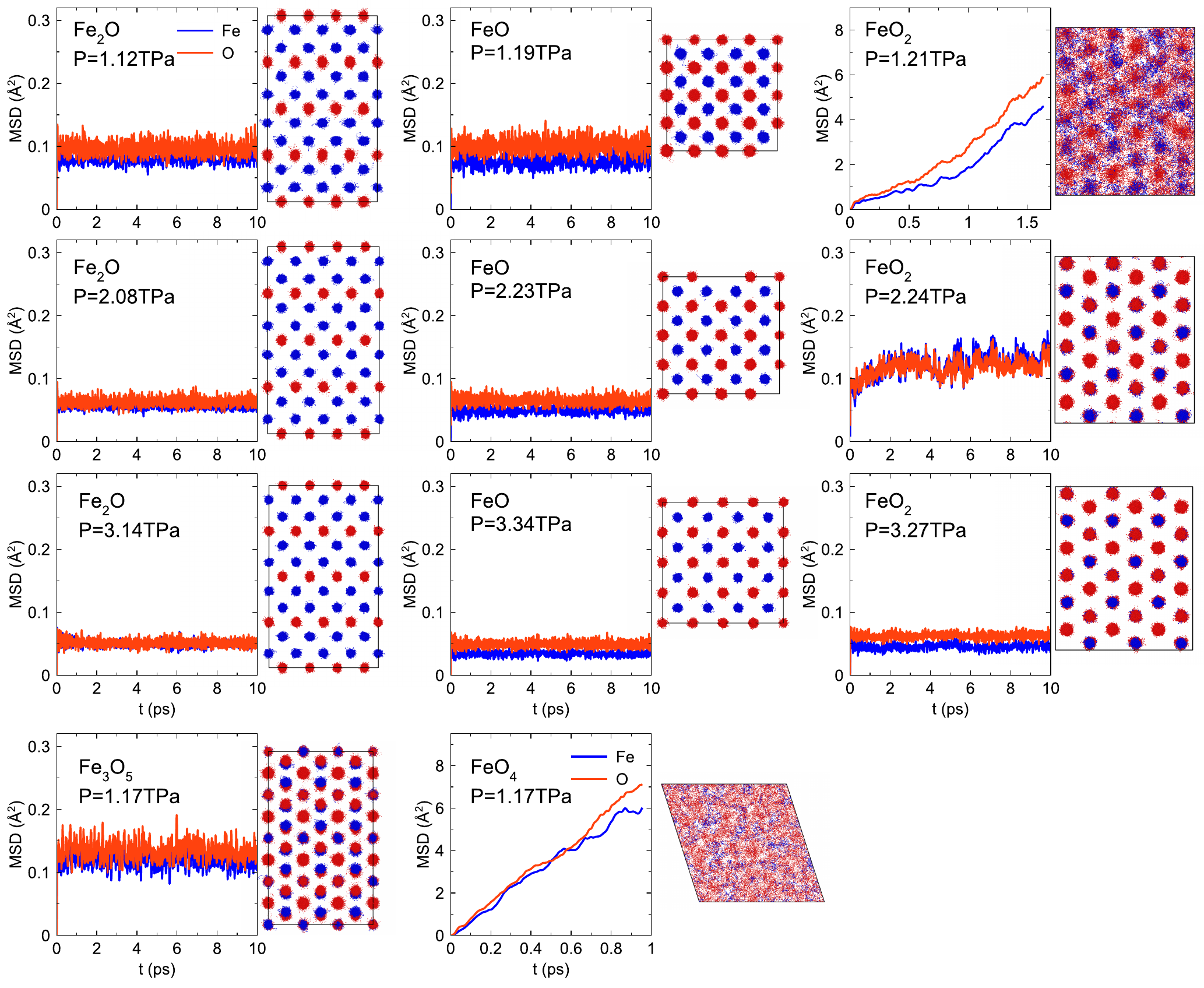}
\end{center}
\caption{Mean square displacement (MSD) of AIMD for ground states at T=8000K. The superposed atomic configurations shown along with the MSD are extracted from AIMD every 5 MD steps. Blue is Fe and red is O. }	
\end{figure}

The electronic density of states (DOS) at the Fermi level (E$_f$) of these five Fe-O compounds is investigated from 1 to 3 TPa. As shown in Fig. 8, our calculations indicate that, except for FeO$_4$, all iron oxides phases identified are metallic. 
It has been reported that the FeO is a typical system that shows insulator to metal transition with increasing pressures. At ambient pressure, FeO shows an insulating B1 phase. When the pressure becomes larger than 120 GPa, the B1 phase transforms to B8 phase, along with an insulator-metal transition and spin transition \cite{a43}. Then the FeO is always metallic at high pressure higher than 150 GPa \cite{a44}. Our current study is in the range of 1-3 TPa, and it is expected that most new FeO phases are metallic. 
For a given Fe$_x$O$_y$ stoichiometry, results indicate that the carrier density decreases with increasing pressure. FeO$_4$ remains an insulating phase up to 3 TPa with band gap reduction, as seen in Fig. S12, despite the PBE/GGA gap underestimation. These results can be attributed to the enhancement of overlap of the atomic wave functions as the pressure increases, which broadens the Fe d bands. As a result, the DOS spreads out in energies at higher pressures in all the crystal phases, as shown in Fig S8-S12. The d band broadening lowers the density of state at E$_f$, i.e. the carrier density. It also suppresses the net moment so that the system favors a non-magnetic state as shown in Fig S1. This mechanism is similar to the pressure effect on the magnet collapse of Fe \cite{a45}. For FeO$_4$, (shown in Fig S12), due to the increase in 3$d$ bandwidth with pressure, the conduction band minimum moves closer to $E_f$ and decreases the band gap. When the pressure is at 3 TPa, the band gap almost vanishes, indicating the FeO$_4$ almost becomes a metallic state. Overall, with the increase of pressure, the broadening of d bands promotes the transition from insulator to metal in FeO$_4$ and decreases carrier density for the metallic states with higher Fe contents.

\subsection{Analysis of structure motifs of Fe-O system under pressure} Besides the stable Fe-O compounds, we also predicted several metastable structures in the Fe-O system from 1 to 3TPa. Since the current calculation does not consider temperature effects on structural stability, these low enthalpy metastable iron oxides may become stable at finite temperatures. Therefore, it is necessary to investigate structural motifs of these stable and metastable FexOy phases to reveal overall structural features in the Fe-O system at high pressures. Here, the threshold for metastability is set to their relative enthalpies (H$_d$) w.r.t the convex hull by 0.7 eV/atom (~ 8000 K). The Fe-centered clusters in these iron oxides were defined by using the cluster alignment method~\cite{36sun2016crystal}. Four typical motifs, including BCC, BCT (body-centered tetragonal), FCC, and HCP are used as templates. We also include the "161" motif (two face-shared hexagonal caps), which is a common cluster in Fe-O binary compounds at high pressure~\cite{14weerasinghe2015computational}.

Snapshots of these motifs are shown in Fig. 9(a). We define an "alignment score" to quantify the similarity between aligned clusters and template motifs~\cite{36sun2016crystal}. Here, the cutoff value of the alignment score is set to 0.125. If the alignment score is higher than 0.125, the cluster is marked as 'others', meaning the group of atoms cannot be classified into the current templates or is much more distorted than these perfect motifs.

\begin{figure}[t]
\begin{center}
\includegraphics[width=2.8in]{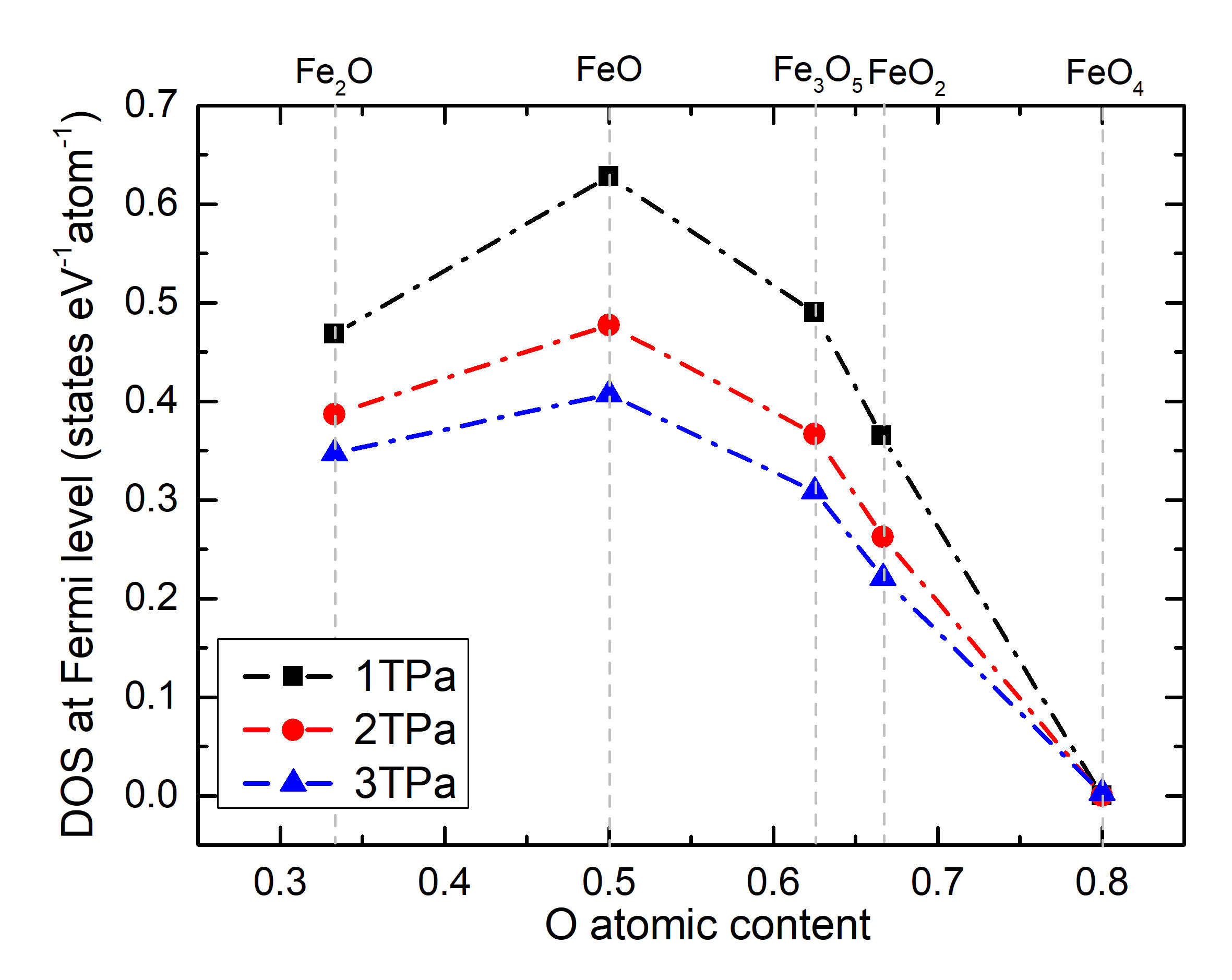}
\end{center}
\caption{(a) Pressure variation of the carrier density in Fe-O compunds. }	
\end{figure}

Fig. 9(b)-(d) shows the relative enthalpies w.r.t. the convex hull of these stable and metastable iron oxides vs. their volumes. The types of Fe-centered clusters and O concentrations are denoted with different symbols and colors, respectively. It can be found that, at 1, 2, and 3 TPa, several motifs may co-exist in Fe-O compound structures with low O content, while, for iron oxides with higher O content, most of them adopt simple BCC motifs, as seen in Fig. 8(b)-(d). Furthermore, we note that several motifs in iron oxides are determined as 'others' at 1, 2, and 3 TPa. Some of them may form more complex clusters than the considered templates, some just highly distorted template-like clusters, e.g., the ground-state $Pnma$ FeO$_2$ structure (Fig. 5(a)).

\section{CONCLUSION}
In summary, we use the AGA method to study structure in the Fe-O system across a wide range of stoichiometries at 1, 2, and 3 TPa. Several stable phases with stoichiometries Fe$_2$O, FeO, Fe$_3$O$_5$, FeO$_2$, and FeO$_4$ are identified. The \textit{ab initio} molecular dynamics simulations indicate that Fe$_2$O, FeO phases have high melting points, larger than 8000 K at 1-3TPa. Therefore, these crystal phases can be stable under the super-Earth's interior conditions. Fe3O5 is stable at 8000K and 1TPa. FeO$_2$ is melt at 8000K and 1TPa while remains stable at 8000K and 3TPa, indicating the melting point increases as a result of increasing pressure. Except for FeO$_4$, the calculated electronic density of states show these Fe-O compounds are metallic. As expected, the carrier density decreases with the increasing pressure. This is due to the enhancement of overlap of the atomic wave functions at high pressure, which broadens the Fe d bands and lowers the density of states at E$_f$, i.e. the carrier density. The cluster alignment analysis reveals that most low-enthalpy phases prefer a BCC packing motif at high pressure, especially those with high O content. This study provides the structural database for the Fe-O system at ultra-high pressure. To fully understand planetary interiors, the joint solubility of high-abundance elements like Fe, Mg, O and Si, etc., under high-pressure conditions must be addressed. For this purpose, the structural behavior of the binary, ternary, etc., systems need to be investigated first. Our study provides necessary information on Fe-O for developing the Fe-Mg-O phase diagram (also essential for the quaternary phase diagram of Fe-Mg-Si-O). It is a preliminary step toward a better understanding of planetary interiors.

\begin{figure}
\begin{center}
\includegraphics[width=3.6in]{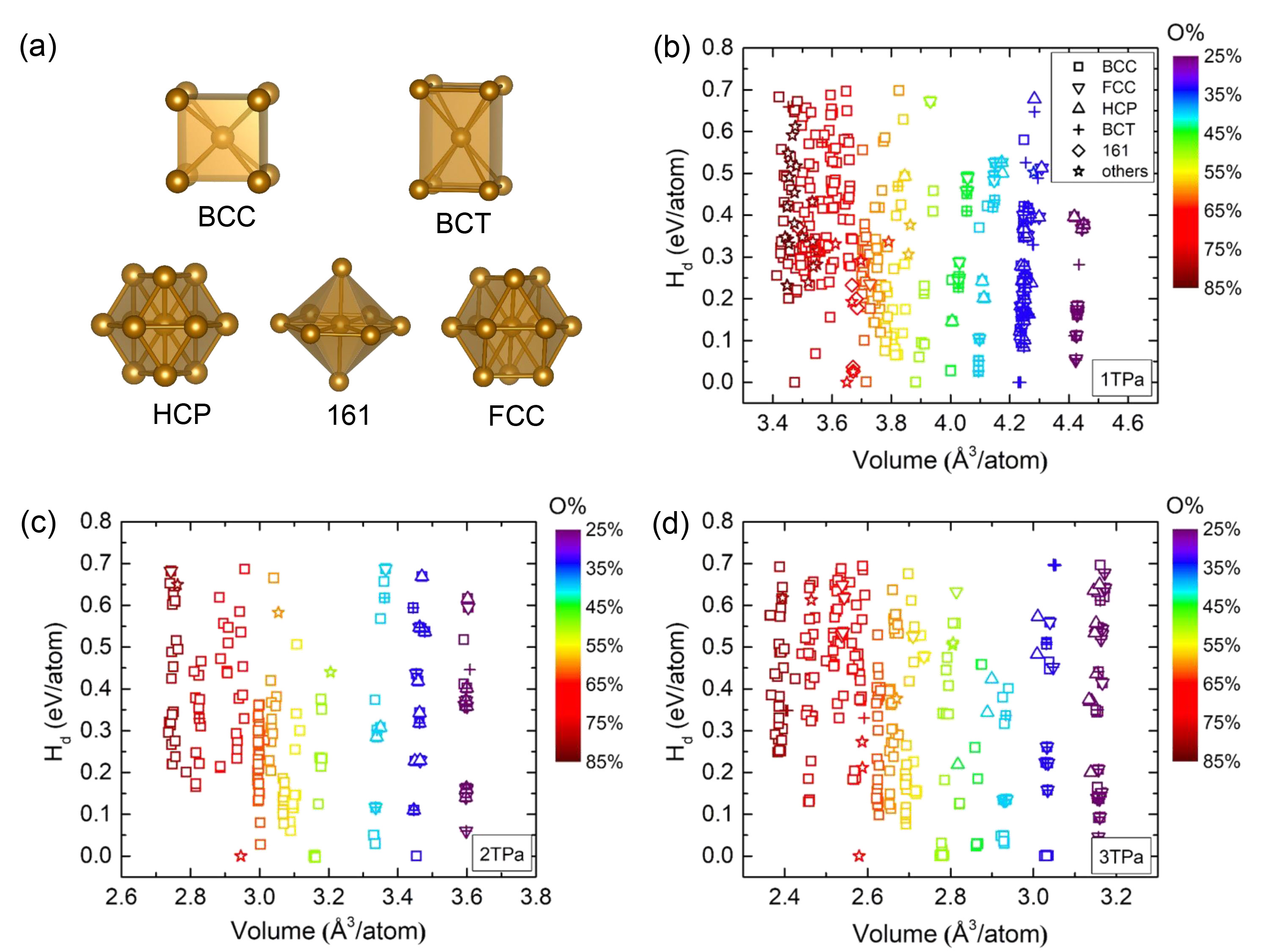}
\end{center}
\caption{(a) The template motifs considered here. (b-d) Enthalpies above the convex-hull (H$_d$) of low-enthalpy Fe$_x$O$_y$ structures v.s. their volumes at 1 TPa, 2 TPa, and 3 TPa. The symbols denote the local packing motifs, and colors represent oxygen concentration.  }	
\end{figure}

\begin{acknowledgments}
Work at Xiamen University was supported by the National Natural Science Foundation of China (11874307). Work at Iowa State University and Columbia University was supported by the National Science Foundation awards EAR-1918134 and EAR-1918126. We acknowledge the computer resources from the Extreme Science and Engineering Discovery Environment (XSEDE), which is supported by the National Science Foundation grant number ACI-1548562. B. D. is supported by JSPS KAKENHI Grant Number JP21K14656. Molecular dynamics simulations were supported by the Numerical Materials Simulator supercomputer at the National Institute for Materials Science (NIMS).
\end{acknowledgments}


\end{document}